\documentclass{article}

\usepackage{amssymb}
\usepackage{amsmath}
\usepackage{booktabs}
\usepackage{graphicx}
\usepackage{float}
\usepackage{subfigure}
\usepackage[hyphens]{url}
\usepackage{mathtools}
\usepackage{geometry}
\usepackage{color}
\definecolor{keynoteBlue}{RGB}{0,127,255}
\definecolor{keynoteYellow}{RGB}{200,127,0}
\definecolor{keynoteRed}{RGB}{178,0,0}
\begin{document}

\begin{titlepage}

\newgeometry{top=2.5cm, bottom=5mm}
\noindent
\large{Title:}\\
\Large\textbf{Assessment of sound spatialisation algorithms for sonic rendering with headsets}\\

\noindent
\large{Authors:}\\
\Large\textbf{
Ali Tarzan\\
RWTH Aachen University\\
Schinkelstr. 2,\\
52062 Aachen\\
Germany\\
telephone: +49 241 8099141\\
email: ali.tarzan@rwth-aachen.de\\
\vspace{0.5cm}\\
Marco Alunno (corresponding author)\\
Universidad EAFIT\\
Cr. 49 \#7sur-50\\
Medellín\\
Colombia\\
telephone: +57 4 2619500 ext. 9102\\
email: malunno@eafit.edu.co\\
\vspace{0.5cm}\\
Paolo Bientinesi\\
RWTH Aachen University\\
Schinkelstr. 2,\\
52062 Aachen\\
Germany\\
telephone: +49 241 8099134\\
email: pauldj@aices.rwth-aachen.de\\
\vspace{0.5cm}\\
This research was conducted at RWTH Aachen University
}

\restoregeometry 
\vfill
\end{titlepage}
\large
\noindent

\section*{Acknowledgment}
Deutsche Forschungsgemeinschaft (DFG) is gratefully acknowledged for making this research possible.
\section*{Funding}
Financial support from the Deutsche Forschungsgemeinschaft (DFG) through grant GSC 111.
\section*{Abstract}
Given an input sound signal and a target virtual sound source, sound spatialisation algorithms manipulate the signal so that a listener perceives it as though it were emitted from the target source. There exist several established spatialisation approaches that deliver satisfactory results when loudspeakers are used to playback the manipulated signal. As headphones have a number of desirable characteristics over loudspeakers, such as portability, isolation from the surrounding environment, cost and ease of use, it is interesting to explore how a sense of acoustic space can be conveyed through them. This article first surveys traditional spatialisation approaches intended for loudspeakers, and then reviews them with regard to their adaptability to headphones.\\

\noindent\textbf{Keywords}: headphones, algorithms, ambisonics, HRTF, spatialisation

\clearpage
\section*{Introduction}
The usage of headphones brings a number of advantages over the usage of loudspeakers, the most obvious ones being the portability headphones offer and the smaller space they take. Loudspeakers usually require a correct setup to work optimally, i.e. specific relative angles and distances to each other. Due to these constraints, the listener is usually expected to be located at a so-called \emph{sweet spot} outside which the system will not perform properly. The listener's orientation is also relevant and head movements often lead to unwanted changes in how sound is perceived. Furthermore, the characteristics of the environment can influence the listening experience significantly when using loudspeakers, since reflections and reverberation affect the way sounds are heard. These issues are non-existent when using headphones: sound arrives directly at the ears of the user and is the same regardless of the listener's position, orientation or environment. This yields other beneficial consequences. In fact, besides safeguarding the privacy of the user, the isolated delivery of sound to one prevents others from being disturbed by undesired acoustic contamination, e.g. when multiple users in the same room use the same application but need to be delivered different sounds, as it happened in multiplayer gaming environments where what one hears strongly depends on a his/her virtual location and orientation. With these advantages of headphones over loudspeakers in mind, it makes sense to explore how well headphones respond to already established spatialisation algorithms. \par
In the present article, we will be discussing spatialisation with ordinary stereo headphones. Although so-called surround sound headphones are commercially available, they are not as nearly common as the others. Also, in order to work properly, they need extra hardware (i.e. digital sound processors with built-in features) to connect to. Therefore, these kinds of headphones will not be taken under consideration.\par 
In Section 1, different spatialisation techniques conceived to be used with loudspeakers are illustrated and discussed. Section 2 reviews each of the previously presented spatialisation techniques and evaluates them with regard to their compatibility with headphones. Since a simple replacement of loudspeakers with headphones is not possible in most cases, due to the high number of output channels used for loudspeaker systems compared to the two channels of conventional stereo headphones, approaches to adapt the existing spatialisation algorithms are examined. Some of them have already been explored, thus current techniques are explained. Other spatialisation algorithms, instead, have apparently not been considered yet. In these cases, possible ideas based on other adaptation techniques are introduced and potential obstacles that hinder their effective application to headphones are pointed out.

\section{Spatialisation approaches}
Sound spatialisation refers to the process of manipulating sound in a way that the listener is able to localise a virtual sound source at a desired location. This virtual sound source does not usually share its position with any of the actual physical sound sources, whether they are loudspeakers or headphones. Ideally, the sound is shaped in a way that the listener does not perceive the existence of the physical sound source at all, but is under the impression that the virtual source is the only source responsible for it. Several well researched and established approaches to spatialisation are presented below.

\subsection{Channel-based systems}
\emph{Stereo and pair-wise mixing.} \label{stereo_pairwise}
The stereo format consists of two independent audio channels usually denoted as \emph{Left} and \emph{Right} channel. Playback is performed by two loudspeakers, one for each channel. Ideally, the loudspeakers face the listener and are positioned at the corners of an equilateral triangle (see Figure 1).

\begin{figure}[htb]
	\centering
  \includegraphics[width=0.5\textwidth ]{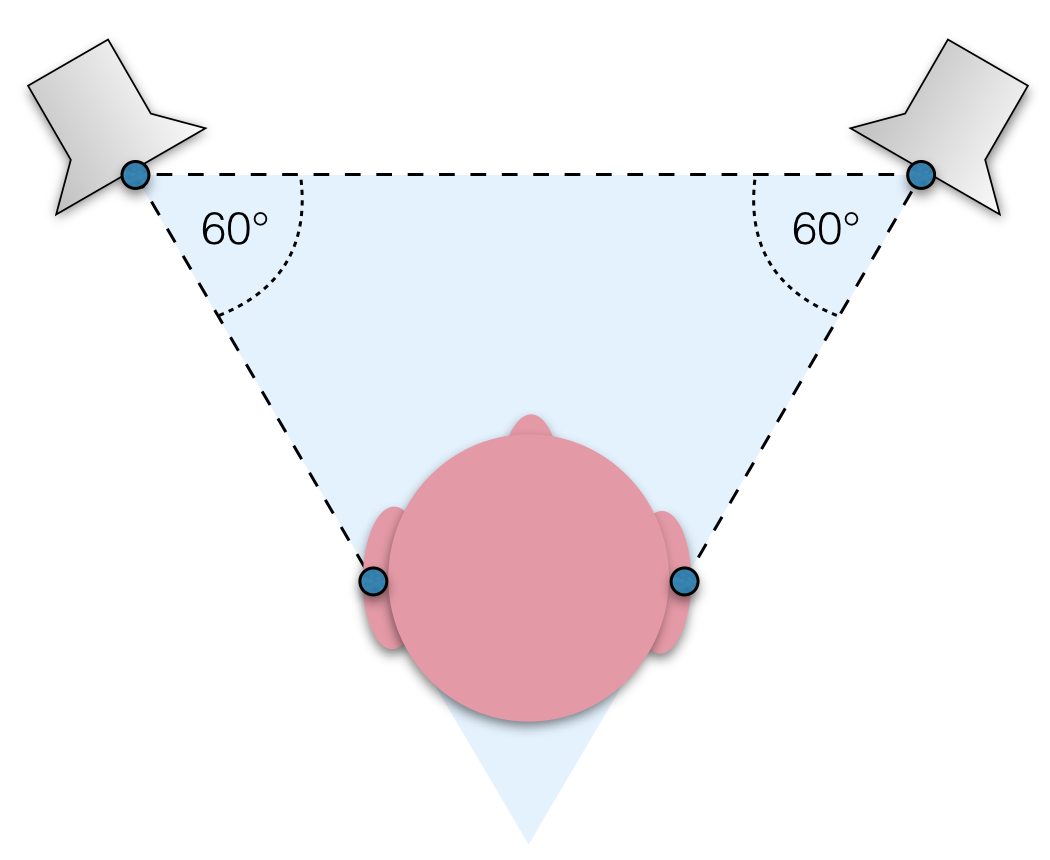}
	\caption{The ideal stereo setup. Loudspeakers are placed and oriented along an equilateral triangle. The listener is positioned such that the ears lie on this triangle.}	
	\label{fig_stereo}
\end{figure}

Stereophonic images can be produced by distributing a given sound signal among the stereo channels such that a listener is able to localise a virtual source that is not perceived at either of the loudspeakers' positions. Of course, a sound that is contained in only one channel and played through only one loudspeaker will be localised by the listener at the position of the respective loudspeaker. However, playing back an identical sound simultaneously through both loudspeakers result in the listener perceiving the source in between the two loudspeakers. In fact, the listener's ears receive two copies of the same sound at the same time (with Interaural Time Difference or ITD = 0) and with identical intensities (with Interaural Level Difference or ILD = 0), which yields a single sound source located in between the loudspeakers, rather than two sources located at the loudspeakers' positions. This virtual sound source that is perceived by the brain at a certain location without being physically there is called a \emph{phantom image} or \emph{phantom source}.

Instead of playing back identical sounds simultaneously to generate a phantom source in between the loudspeakers, both ILD and ITD can be manipulated by adding a slight delay or changing the loudness to one of the channels. With this approach called \emph{pair-wise mixing} or (\emph{pair-wise panning}), a phantom source can be generated at any position on the line linking the loudspeakers. The position of the phantom image varies with the delay or amplitude difference. For a delay of approximately 2 ms or more -- and unaltered amplitude -- the phantom source is perceived at the position of the loudspeaker with the non-delayed signal. However, the delay's value before the virtual source is drawn into one of the loudspeakers can be longer for more complex sounds like human speech. 

As shown in Figure 2, for non-delayed signals, an amplitude difference of approximately 16 dB yields a phantom source at the loudspeaker with the higher amplitude (Martin et al., 1999). The position of the virtual source can be controlled with the tangent law (Pulkki \& Karjalainen, 2001):
\begin{equation}
\frac{\tan{\phi_v}}{\tan{\phi_l}} = \frac{g_1 - g_2}{g_1 + g_2}
\end{equation}
where $\phi_v$ is the angle between the listener and the virtual source, $\phi_l$ is the angle between the listener and the loudspeakers ($30^\circ$ in Figure 1) and $g_i \in [0,1]$ are the individual gains.

\begin{figure}[htb]
	\centering
  \includegraphics[width=1\textwidth ]{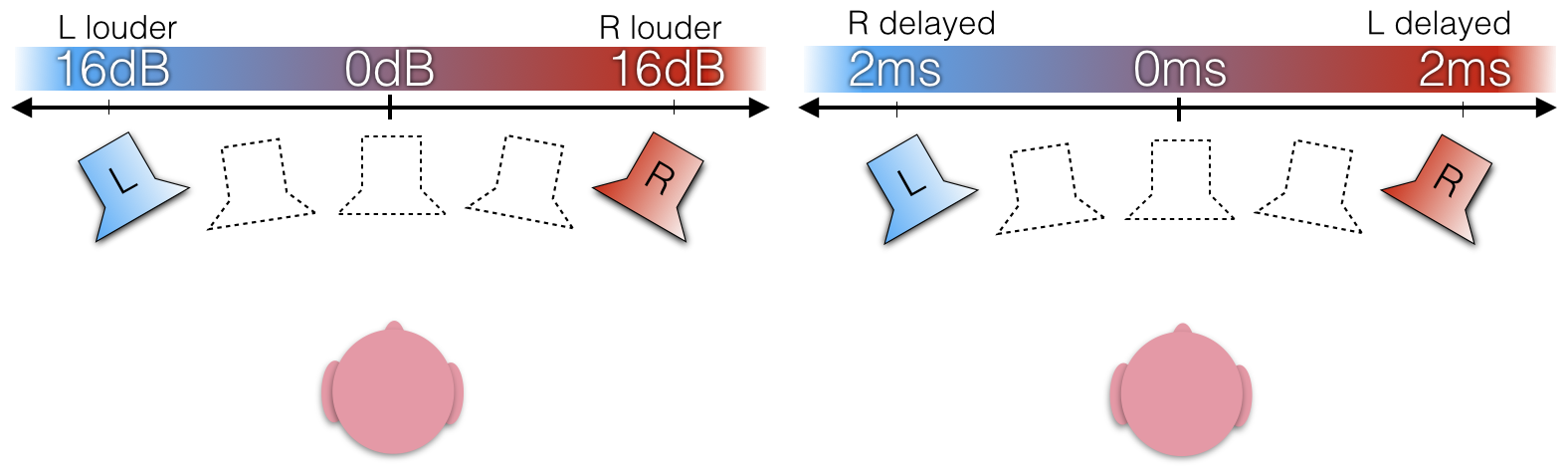}
	\caption{The resulting position of a phantom source. Left: Different amplitudes played at the same time. Right: Same amplitudes with one of the signals delayed.}	
	\label{fig_pairwise}
\end{figure} 

Interestingly, by artificially manipulating binaural cues it is also possible to create contradicting cues that would never be experienced with natural sound sources, e.g. making the sound arrive earlier at the left ear but louder at the right ear. In this case, the perceived location is a compromise of what is suggested by each cue, unless the sound is in a frequency range where one of the binaural cues strongly dominates over the other (Dickreiter et al., 2014; Schnupp et al., 2011).\\

\noindent\emph{Quadraphonic.}
While the pair-wise mixing approach allows for spatialisation of sound along one dimension, the quadraphonic setup is meant to extend this principle to two dimensions. Quadraphonic was one of the earliest surround techniques when it appeared in the 1970s and is often considered to be the predecessor of today's popular $x.1$ surround systems (most notably \emph{5.1}). As the name suggests, the quadraphonic setup consists of four loudspeakers arranged at the corners of a square with the listener at its centre (Figure 3). 

\begin{figure}[htb]
	\centering
  \includegraphics[width=0.34\textwidth ]{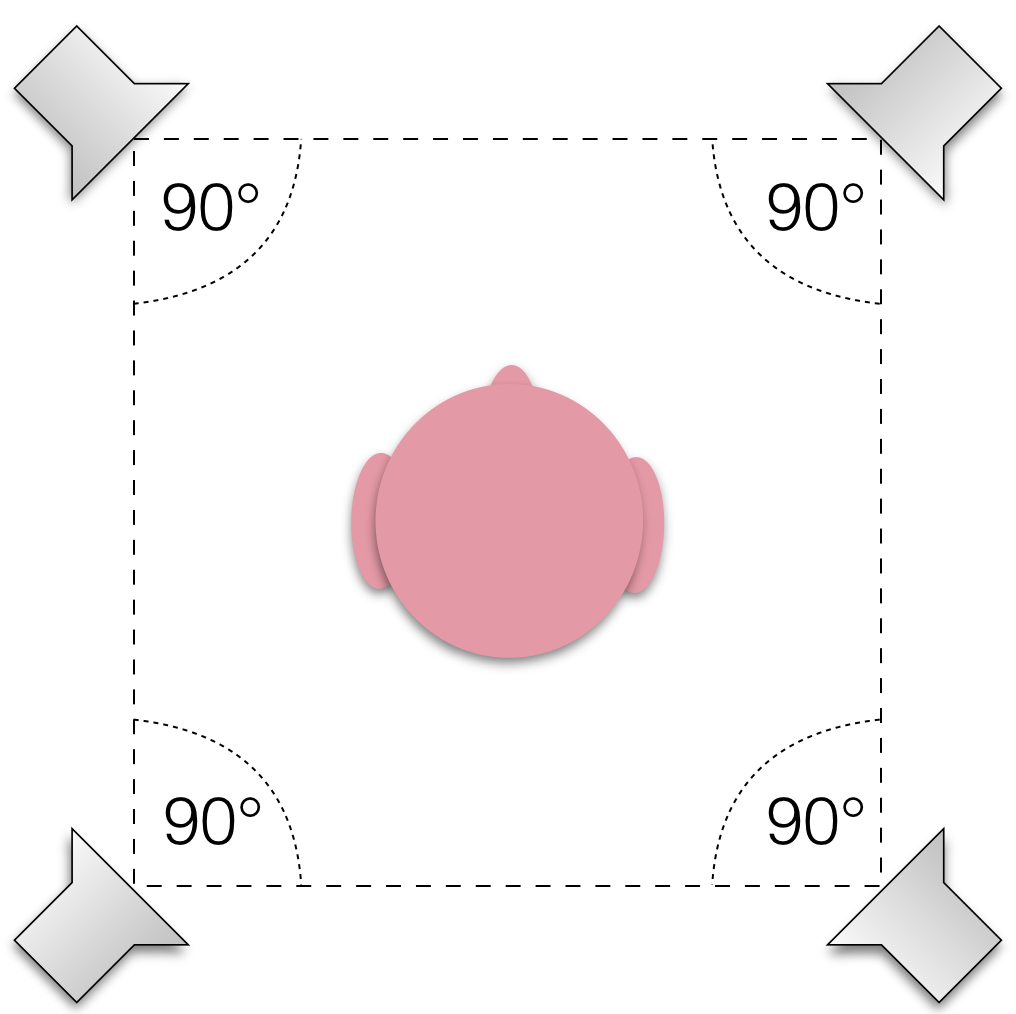}
	\caption{Loudspeaker setup for \emph{quadraphonic}.}	
	\label{fig_quad}
\end{figure}

The initial hope was to be able to spatialize sound in two dimensions by generating phantom images between adjacent loudspeakers through pair-wise mixing. Important to note is that for any given phantom image, only two loudspeakers would be contributing, while the others remain silent (or contribute to other images). While spatialisation for stereo setups works reasonably well as long as loudspeakers are, at most, $60^\circ$ apart and in front of the listener, applying this approach for angles wider than $60^\circ$ turns out to be problematic. The $90^\circ$ angle between adjacent loudspeakers that is used in the quadraphonic setup is too wide for covering the whole area without creating gaps. Thus, phantom images are either drawn to individual loudspeakers or are not formed at all and the listener perceives separated signals coming from different directions. These problems apply to localisation between all four pairs of adjacent loudspeakers. Additionally, for the rear pair, this approach suffers from unstable images as they are strongly displaced when the listener's head is slightly tilted or moved. Also, it is hard to place virtual sources between the loudspeakers since phantom images are drawn to the non-delayed loudspeaker for very short delays (or to the higher amplitude loudspeaker for slight differences of amplitude). The worst situation occurs when trying to create phantom images at either side of the listener, as it has been demonstrated by multiple studies (Elen, 2001; Gerzon, 1985; Theile, 1977).\\

\noindent\emph{x.1.}
Today's surround systems have additional channels compared to quadraphonic. To counter the gap caused by the large separation angle of $90^\circ$ between the front-left and front-right loudspeakers, an additional loudspeaker is usually placed in between. The most common setup is the \emph{5.1} and consists of five loudspeakers (arranged as in Figure 4) and an additional channel for low-frequency sounds. The position of the fifth loudspeaker (subwoofer) is not relevant, since humans are quite insensitive to localisation for low frequencies.

\begin{figure}[htb]
	\centering
  \includegraphics[width=0.45\textwidth ]{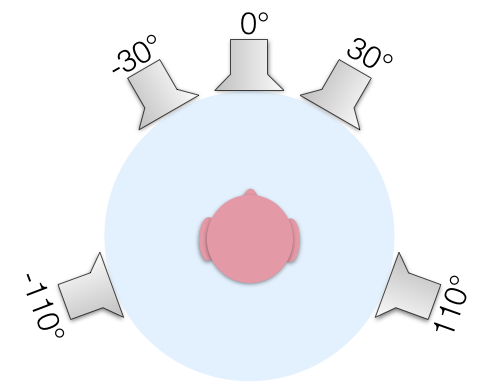}
	\caption{Five loudspeaker setup as recommended by the ITU (2006). All loudspeakers have the same distance to the listener.}	
	\label{fig_itu51}
\end{figure}

This setup solves the issue of poor localisation at the front observed in the quadraphonic setup, but still suffers from bad localisation between rear and side channels. Therefore, rear channels are often used for ambient sounds (e.g. rain) that do not require any specific localisation while still giving the experience of a surrounding soundscape.
Figure 5 shows a summary and comparison of the channel-based approaches

\begin{figure}[htb]
	\centering
  \includegraphics[width=1\textwidth ]{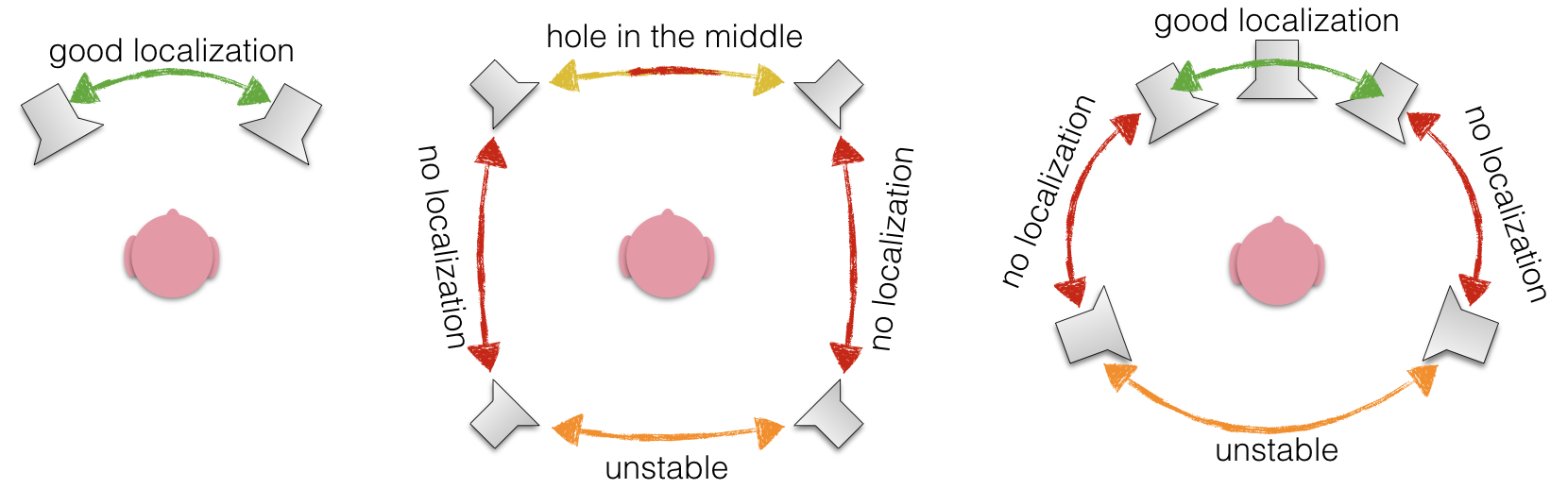}
	\caption{Comparison of localisation performance of channel-based systems when using the pair-wise mixing approach. From left to right: Stereo, Quadraphonic and 5.1.}
	\label{fig_localCompa}
\end{figure}

Additional channels can be added to extend the \emph{5.1} setup: the \emph{7.1} setup adds two loudspeakers on the side and the \emph{11.1} introduces elevated loudspeakers in an attempt to create three-dimensional acoustic environments. Other multichannel formats exist, a notable example being the \emph{octophonic} system where eight loudspeakers are arranged either at the corners of a cube for three-dimensional sound or equally spaced on a circle around the listener for two-dimensional sound.   

\subsection{Ambisonics}
Ambisonics is a surround sound technique developed in the 1970s, most notably by Michael A. Gerzon and Peter Fellgett at the National Research Development Corporation. Despite its lack of commercial success, it has several significant advantages over better known surround sound systems such as \emph{5.1}. One of these advantages is that Ambisonics can be used to deliver a full-sphere surround sound, as opposed to the two-dimensional sound offered by traditional surround systems. Another advantage is the flexibility of the loudspeakers setup: while other surround systems strictly dictate the number of loudspeakers that can be used and their positions, with Ambisonics these parameters are open for individual customisation by the end user. Further, Ambisonics overcomes the problem of poor localisation from certain directions. \\

\noindent\emph{The B-format.}
A basic format in which Ambisonics can be stored and distributed is the \emph{B-format}. The \emph{B-format} consists of four signals: \emph{X, Y, Z} and \emph{W}. \emph{W} contains global sound pressure information while \emph{X, Y} and \emph{Z} contain directional information for each of the three dimensions of space. In a two-dimensional case, the signal \emph{Z} is always $0$. It is important to note that these four signals do not correspond to four loudspeakers' channels in any way. Unlike traditional surround systems where each channel of the data format contributes to the output of one loudspeaker, here the resulting output of each loudspeaker is computed by a decoder using all four signals of the \emph{B-format}. A related feature of Ambisonics is that the content of the \emph{B-format} does not specify how many loudspeakers have to be used for playback.

There are two ways to fill the \emph{B-format} with data: one way is to record the components of the \emph{B-format} directly with a suitable microphones setup, another way is to take monophonic input signals and placing them in the three-dimensional space by encoding them into the four aforementioned components of the \emph{B-format}. Either way, after the \emph{B-format} has been obtained, a special Ambisonic decoder that contains all the information about the loudspeakers setup is needed to playback the result. \\

\noindent\emph{Encoding the B-format.}
\emph{Encoding} or \emph{panning} a sound source into the B-format does not require any specific Ambisonic recording setup. \emph{Panning} refers to the process of taking a monophonic sound signal and placing it in a desired target direction in the three-dimensional space. By playing back the resulting B-format file with a decoder, the listener is able to hear the sound as coming from the target location. The \emph{encoder} receives the monophonic signal as an input $I$ as well as a horizontal angle $\theta$ and an elevation angle $\phi$ (in the two-dimensional domain $\phi$ can be omitted or set to $0$). The panner then takes care to distribute the input signal $I$ among the components of the B-format according to the following equations (Gerzon \& Barton, 1984; Malham, 1998):
\begin{equation} \label{eqAmbiPanner}
W = \frac{I}{\sqrt{2}}
\end{equation}
\begin{equation}
X = I \cdot \cos{\theta} \cdot \cos{\phi}
\end{equation}
\begin{equation}
Y = I \cdot \sin{\theta} \cdot \cos{\phi}
\end{equation}
\begin{equation} \label{eqAmbiPannerEnd}
Z = I \cdot \sin{\phi}
\end{equation} 
Note that $W$, that contains global sound pressure levels, does not depend on $\theta$ or $\phi$, but it is multiplied by a scalar so that the average energy levels of all four channels are approximately the same. Alternatively, $\{X, Y, Z\}$ could be multiplied by $\sqrt{2}$ while $W$ receives the unaltered input $I$. The final signals set $\{W, X, Y, Z\}$ is the same set that would be computed by recording a sound source at a specific location with a native B-format microphones' array (Gerzon, 1980).

In order to create an ambisonic soundscape, the encoding process can be done by using multiple input signals with different $\theta$ and $\phi$. A basic encoder allows the user to place a sound signal only in a certain direction via the angle parameters. The distance coordinate that is required to specify a point in space within a spherical coordinate system (or polar coordinates in the two-dimensional case) is not part of this basic version. When a soundscape is created by panning multiple monophonic input signals, their relative sound distances can be mimicked to a certain degree with relative amplitudes, i.e multiplying all components with gain coefficients representing the distance in the above Equations (\ref{eqAmbiPanner})-(\ref{eqAmbiPannerEnd}) (Schacher \& Kocher, 2006). This simple way to model distance through amplitude has its limits, though. In fact, humans use a more complicated set of cues to determine the distance of a sound source, such as the ratio of direct sound to reverb and the fact that higher frequencies fade away faster than lower frequencies (Blauert, 1977; Malham, 1998). More advanced panners are also able to apply near field effects that boost certain frequencies of close sound sources by introducing an additional parameter for the distance (Daniel, 2003). An expansion of the B-format with an additional angular parameter for distance has been proposed in (Penha, 2008). \\

\noindent\emph{Manipulating the B-format.} \label{manipBformat}
Another advantage of Ambisonics is the possibility to edit the soundscape captured in the B-format after recording or encoding it. The soundscape can be rotated around any of the three axes (see Figure 6) with basic matrix operations described in (Malham, 1998). 

\begin{figure}[h]
	\centering
  \includegraphics[width=0.37\textwidth ]{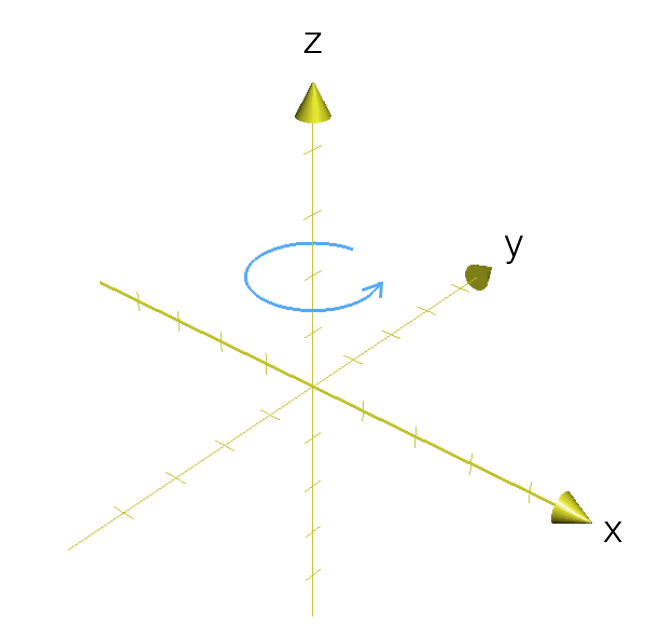}
	\caption{Rotation around the z-axis.} 
	\label{fig_rotation}
\end{figure}

These rotations can be chained to freely move the orientation of the soundscape around. However, while rotating a point around an axis, the distance from the point to the origin of the coordinate system remains the same, thus only the direction from where individual sounds arrive to the listener is affected. More complex operations allow directional loudness modifications, i.e increasing/decreasing the loudness of sound from certain angles (Gerzon \& Burton, 1992; Kronlachner, 2014). \\

\noindent\emph{Decoding the B-format.}
\label{sec_decoding}
Since data in the B-format do not correspond to loudspeakers as it is the case with other surround systems, the B-format needs to be \emph{decoded} before playback. This process computes a linear combination of all signals of the B-format for each loudspeaker:
\begin{equation} \label{eq_decodmatrix}
\begin{bmatrix}
    L_{1}\\
    L_{2}\\
   \vdots\\ 
    L_{n}
\end{bmatrix}
=
 \begin{bmatrix}
   D_{W1} & D_{X1} & D_{Y1} & D_{Z1}\\
   D_{W2} & D_{X2} & D_{Y2} & D_{Z2}\\
   \vdots  & \vdots  & \vdots  & \vdots \\
   D_{Wn} & D_{Xn} & D_{Yn} & D_{Zn}\\
\end{bmatrix}
 \begin{bmatrix}
   W  \\
   X  \\
   Y  \\
   Z  \\
\end{bmatrix}
\end{equation}
where $L_i$ is the output signal of the loudspeaker $i$ and $D$ is the decoder matrix that needs to be found. There are several different approaches to obtain $D$. Usually, while being flexible with regard to the number of loudspeakers and, to a degree, also to their arrangement, a decoder imposes some constraint on the layout of the loudspeakers setup, i.e. the number of loudspeakers must be at least one more than the number of signals used. For example, the two-dimensional case, where only W, X, and Y are used, requires at least four speakers for reasonable playback. Heller et al. (2008) group loudspeakers' layouts in three categories:
\begin{enumerate}
\item regular polygons and polyhedra
\item irregular layouts but with speakers in diametrically opposite pairs
\item generally irregular
\end{enumerate}
Figure 7 shows an illustration of these three groups. Note that the distance to the listener should be the the same for loudspeakers of a diametrically opposing pair. However, any divergence from this scheme can be compensated if the output of the closer loudspeaker is delayed accordingly.

\begin{figure}[h]
	\centering
  \includegraphics[width=0.90\textwidth ]{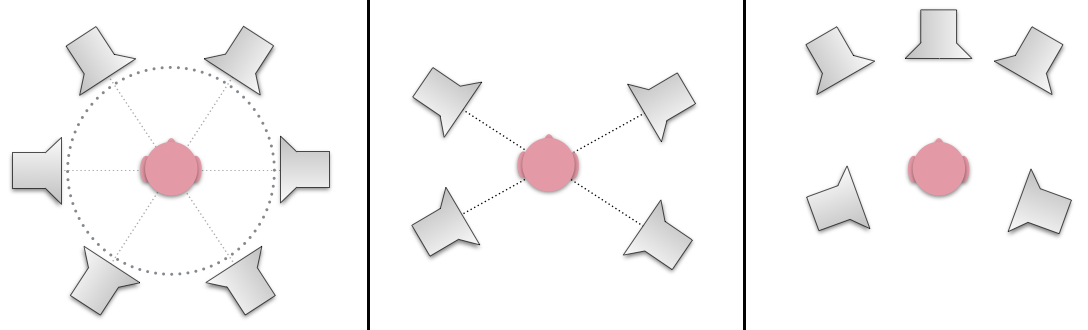}
	\caption[caption]{Examples for each layout category identified in Heller et al. (2008).} 
	\label{fig_decoCatego}
\end{figure}

\subsection{Higher Order Ambisonics}
The Ambisonics approach described so far has already numerous advantages over traditional surround systems. However, it suffers from a relatively small \emph{sweet spot}, that is, the area where a listener can experience an accurate reproduction of the sound field is fairly limited. Moving away from this area gradually decreases localisation quality and this effect becomes stronger for higher frequencies (Bamford \& Vanderkooy, 1995). A solution to this problem as well as an increase of the spatial resolution is offered by \emph{Higher Order Ambisonics}, an extension of Ambisonics. Traditional Ambisonics, as described up to this point, are a special case of Higher Order Ambisonics, namely Higher Order Ambisonics of order 1. As a reminder: the B-format consists of four components  -- W, X, Y and Z -- that play different roles in encoding the location of the sound source. Each of them can be recorded with microphones that have basic polar patterns. These patterns can be described with functions called \emph{spherical harmonics}. An equation that describes spherical harmonics is the following:
\begin{equation} \label{eqSpherical}
Y^\varsigma_{mn}(\theta, \phi) = \tilde P_{mn}(\sin \phi) \cdot \begin{cases}\cos(n\theta) \quad if \quad \varsigma = 1\\\sin(n\theta) \quad if \quad \varsigma = \text{-}1\end{cases}
\end{equation}
where $\tilde P_{mn}$ is the \emph{associated Legendre function} (Abramowitz \& Stegun, 1964) with degree $m \in \mathbb{N}$ and order $n \in \mathbb{N}, n \leq m$. (Malham, 2003). For $m \leq 1$ the resulting spherical harmonics correspond to the already known patterns of order 1. Higher Order Ambisonics with $m>1$ introduces additional signals that are used alongside the signals of first order W, X, Y and Z. Figure 8 shows an illustration of spherical harmonics up to order 3 ($m \leq 3$).

\begin{figure}[h]
	\centering
  \includegraphics[width=0.90\textwidth ]{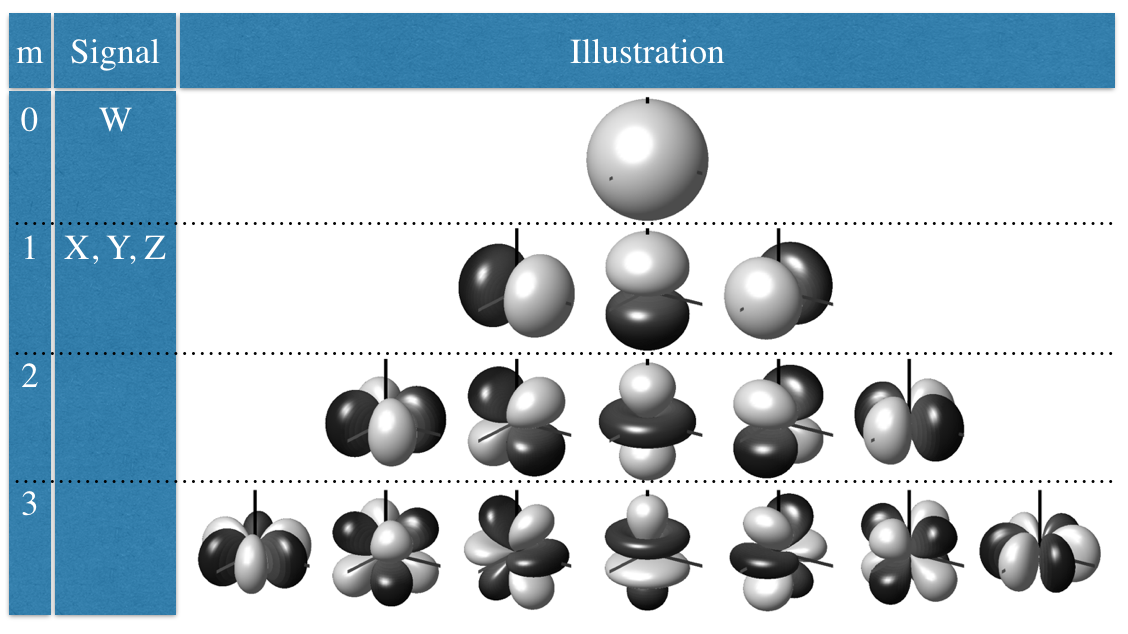}
	\caption[caption]{Illustration of spherical harmonics up to order 3. Note that orders 0 and 1 correspond to the signals of the B-format. (Illustration from Wikimedia Commons by Franz Zotter (CC BY 3.0))} 
	\label{fig_HOA}
\end{figure}

The improvement in spatial resolution and an increased size of the sweet spot comes at the cost of having to use additional signals in comparison to the initial B-format. This goes hand in hand with a higher number of loudspeakers that have to be used for playback (see subsection \textit{Decoding the B-format}). Table 1 shows the number of components needed for a system of a given order. It can be seen that in the full-sphere case the number of signals increases quadratically with the order. Considering the higher sensibility of humans to the horizontal plane compared to the sensibility to vertical cues (Blauert, 1997), this uniformly increased complexity seems partially unnecessary. For this reason, models for \emph{mixed-order Ambisonics} have been proposed. When using mixed-order schemes, the order of the system is no longer uniform for the whole sphere and, thus, no longer defined by a single parameter $P$, the periphonic order. Instead, different orders can be defined for the horizontal and the vertical planes. This is achieved by combining appropriate signal sets, where higher orders components are selected only for the horizontal plane. For example, by taking all signals from the first two rows and only the outer most components in the third row of Figure 8, a mixed-order system with a horizontal order of 2 and a periphonic order of 1 is obtained. A widely known mixed-order scheme is the two parameter scheme \emph{\#H\#P}. The parameter $H$ defines the order in the horizontal plane, whereas the parameter $P$ defines the periphonic order. The \emph{\#H\#P} scheme defines only the components that are used for a given pair of parameters, i.e \emph{\#3\#1} refers to a unique set of components and is not arbitrarily chosen.

\subsection{Wave Field Synthesis}
Wave Field Synthesis (WFS) is a spatialisation technique proposed by A. J. Berkhout in 1988 (Berkhout, 1988). It differs from the other techniques presented above in that it aims to recreate a wavefield in a larger area using physical principles rather than relying on psychoacoustics to deliver the perception of virtual sources to a listener at a specific location. This is achieved by placing the listener inside a large array of loudspeakers that are all individually controlled.

The concept of \emph{wavefront} is essential to understand how WFS works. A wavefront of a wave is a set of points for which it would take the wave the same time to travel to from the wave source. Consequently, all points of the wavefront have the same phase. For sources that emit sound in all directions, the shape of the wavefronts is spherical, while if sound is spread on a plane their shape is circular, as shown in Figure 9.

\begin{figure}[h]
	\centering
  \includegraphics[width=0.6\textwidth ]{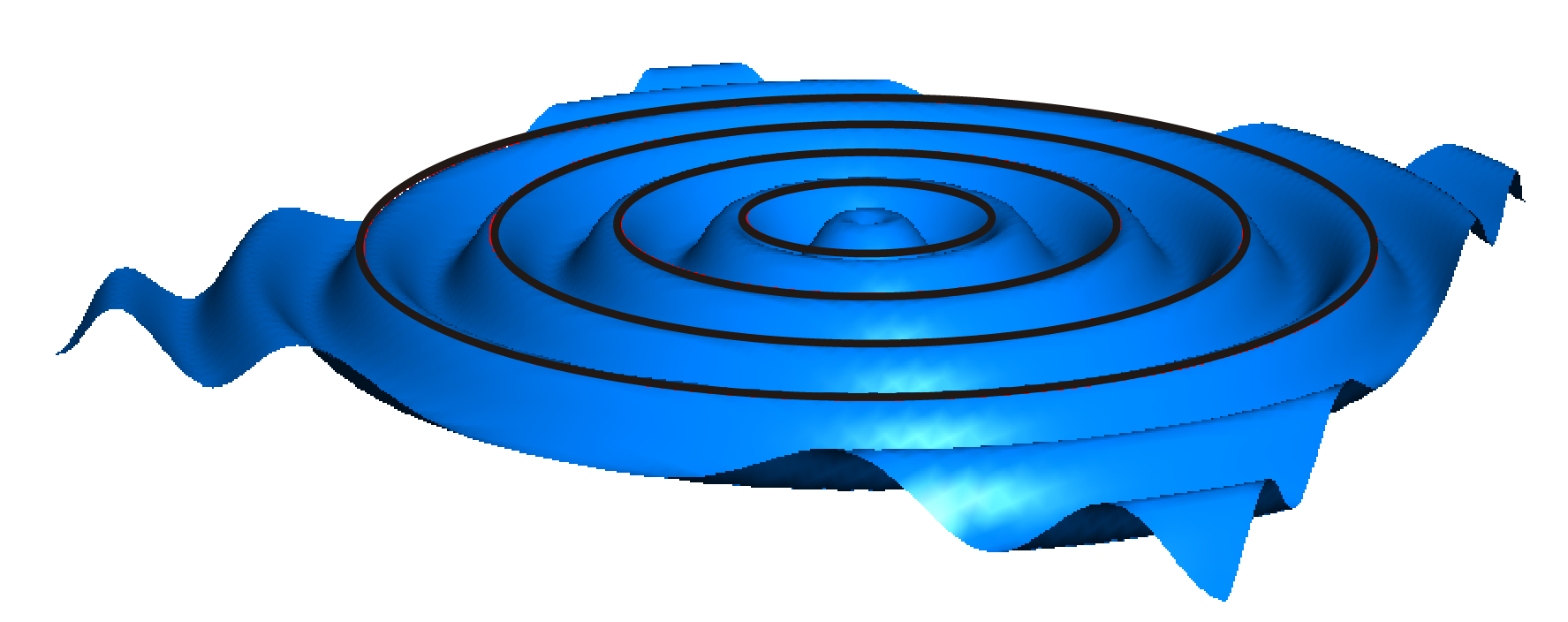}
	\caption[caption]{Visualisation of wavefronts (black) of a two dimensional wave. The wave is emitted in all directions. Any circle around the source is a wavefront.}
	\label{fig_wavefronts}
\end{figure}

\begin{figure}[h]
	\centering
  \includegraphics[width=0.7\textwidth ]{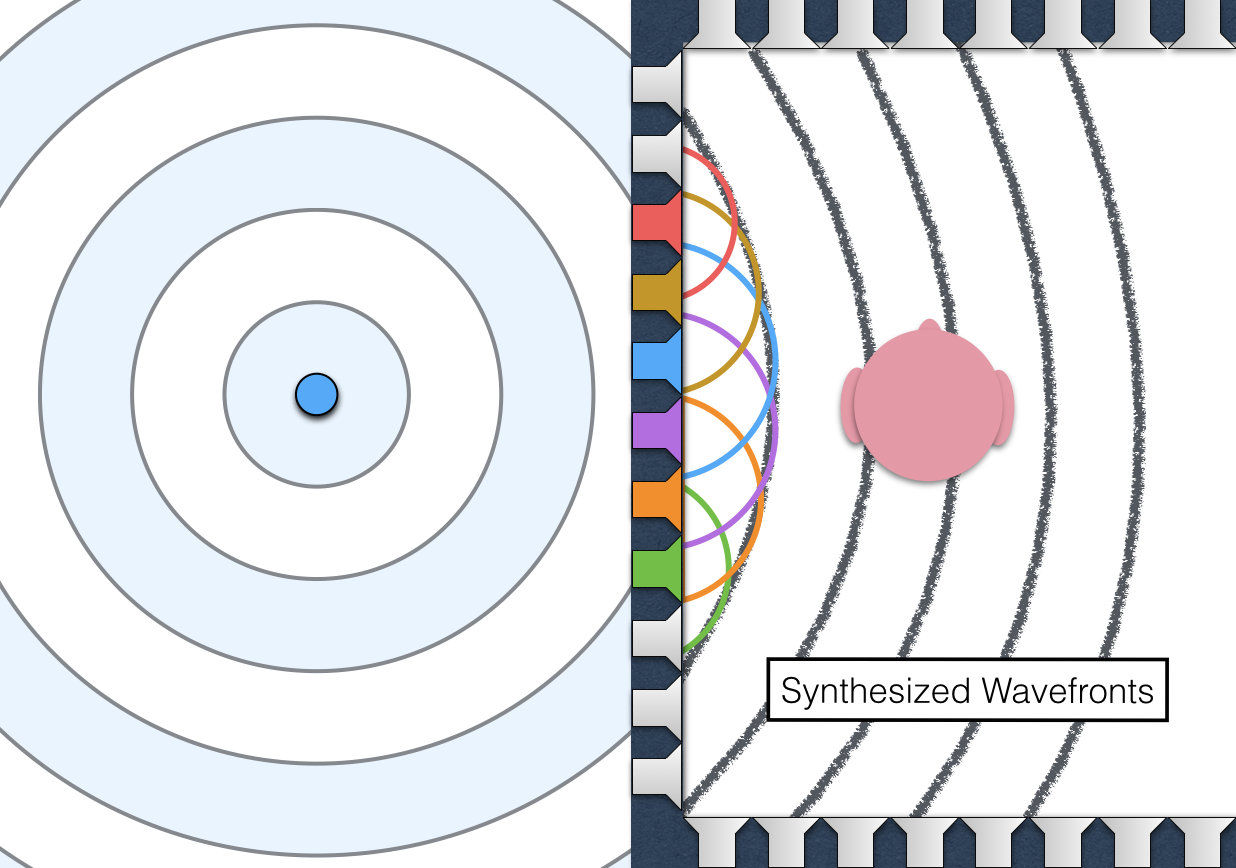}
	\caption[caption]{The principle of Wave Field Synthesis illustrated. The loudspeakers use elementary waves to synthesise wavefronts. The listener perceives these synthesised wavefronts as if they were created by the virtual source on the left.} 
	\label{fig_wfs}
\end{figure}

The physical principle that makes Wave Field Synthesis possible is the \emph{Huygens-Fresnel principle}. It states that every wavefront can be decomposed into a set of spherical waves called elementary waves. Conversely, any possible wavefront can be synthesised by elementary waves. These elementary waves are created by the loudspeakers. The interference of these elementary waves create artificial wave fronts that are nearly identical to wavefronts created by real sound sources. This principle is illustrated in Figure \ref{fig_wfs}.

The wave equation is a partial differential equation that allows to describe the characteristics of a wave. With the given pressure levels at the loudspeaker positions as a boundary condition, the \emph{Kirchhoff-Helmholtz integral} (Williams, 1999) is the solution to the wave equation. It allows to compute the sound pressure level in any point of a bounded region, provided the pressure levels at all points of the boundary region (or `surface') are known and the region is source-free, i.e. there are no sources inside of the region. The acoustics of the room, e.g. reflections against the walls, may prevent the region from being completely source-free, but placing the loudspeakers setup in an anechoic room can minimise this effect. Alternatively, the influence of room acoustics on the listener can be reduced if the listener is located in the near-field of the loudspeakers.

While the principle of WFS works for three dimensions, the actual realization of a three-dimensional setup is barely feasible due to the high number of loudspeakers required. Therefore, practical applications use a restricted version where sound is reproduced on a two-dimensional plane. However, virtual sources can be better modelled if the arrays of loudspeakers are located at ear-level, which means that virtual sound sources elevated above or below the height of the loudspeakers can no longer be considered. Additionally, listeners whose ears are not located on the same plane as the loudspeakers will hear artifacts that may lead to a wrong localisation of the virtual sound sources. Aside from these restrictions, the two-dimensional case allows for highly accurate reproduction of wave fields in a plane if loudspeakers are not more than about 4-6 inches apart. In fact, a bigger distance between loudspeakers yields audible aliasing effects (Rabenstein \& Spors, 2006).

WFS has a big advantage over other spatialisation techniques: by reconstructing the wave field in a large area, its performance works properly independently on the listener's position and orientation, as long as the listener moves inside the volume enclosed by the loudspeakers' array. This makes WFS very useful for applications where the users need to move around freely, e.g. in Virtual Reality environments. Moreover, spatialisation can be experienced by multiple users at the same time.

The main weakness of Wave field Synthesis is its cost and the complexity of the setup. Even in a two-dimensional case, it uses significantly more loudspeakers than other techniques. Additionally, a large anechoic room is required.

\subsection{Vector Base Amplitude Panning}
As the name suggests, \emph{Vector Base Amplitude Panning (VBAP)} is a method that uses amplitude panning to position virtual sound sources around the listener. It works on a full sphere as well as on the horizontal plane only and with any number of loudspeakers, as long as they are equidistant from the listener. The virtual sound sources can be stationary or moving and many of them are allowed to be active at the same time. 

It works by selecting a subset of loudspeakers and computing individual gains for each loudspeaker so that phantom images are generated. In the horizontal case, only two loudspeakers are selected, making this condition similar to the pairwise-mixing approach described in Section \ref{stereo_pairwise}. In the three-dimensional case, three loudspeakers are selected and the virtual source can be placed anywhere in the triangle formed by the selected loudspeakers.  

The first step is to define a set of bases from the set of loudspeakers. A base is a pair of loudspeakers in the two-dimensional and a triplet in the three-dimensional case. Each virtual source will be generated by one base (but several virtual sources can belong to the same base at the same time). Since virtual sources will always be located inside the area enclosed by the loudspeakers used to generate them, the maximal error is limited by the distance of the loudspeakers inside a base. Therefore,  bases are ideally formed by loudspeakers that are close to each other, i.e. adjacent loudspeakers. Also it is advantageous for moving sources if the regions covered by different bases do not overlap. Figure 11 shows a set of bases and their respective active area.

\begin{figure}[h] 
	\centering
  \includegraphics[width=0.5\textwidth ]{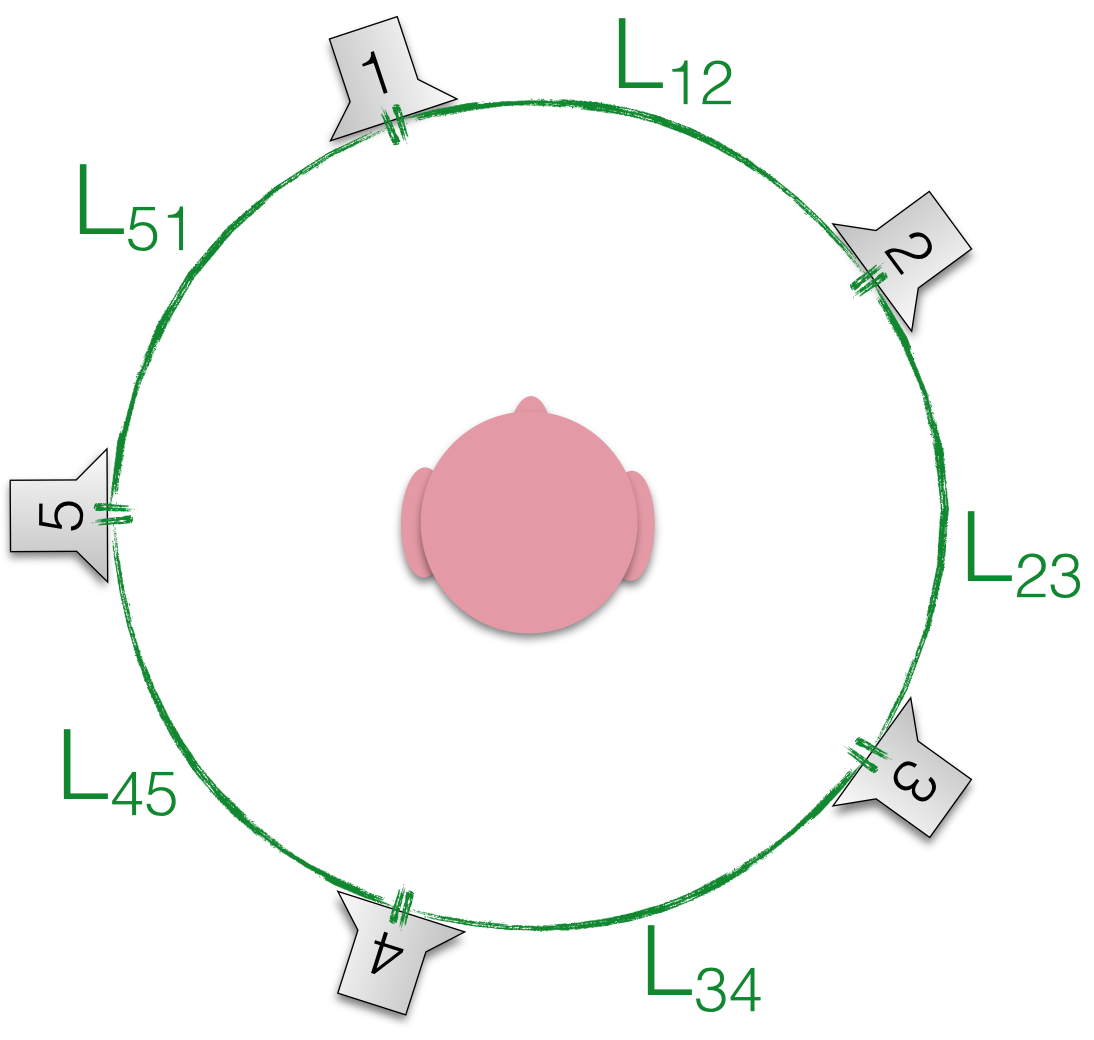}
	\caption[caption]{The bases $L_{ik}$ are chosen to be formed by adjacent loudspeakers.} 
	\label{fig_vbap_bases}
\end{figure}

After the set of bases is defined, a virtual source is generated by first selecting a base whose active area contains the position of the virtual source. (The computational method for finding the base will be presented later. As of now, let us assume that the base is already selected). Each loudspeaker $i$ has a unit vector $l_i$ pointing from the listener to the loudspeaker. The goal is to express the unit vector $p$ pointing from the listener to the virtual source through a linear combination of loudspeakers' vectors $l_i$ and respective gains $g_i$ (see Figure 12):

\begin{figure}[h] 
	\centering
  \includegraphics[width=0.75\textwidth ]{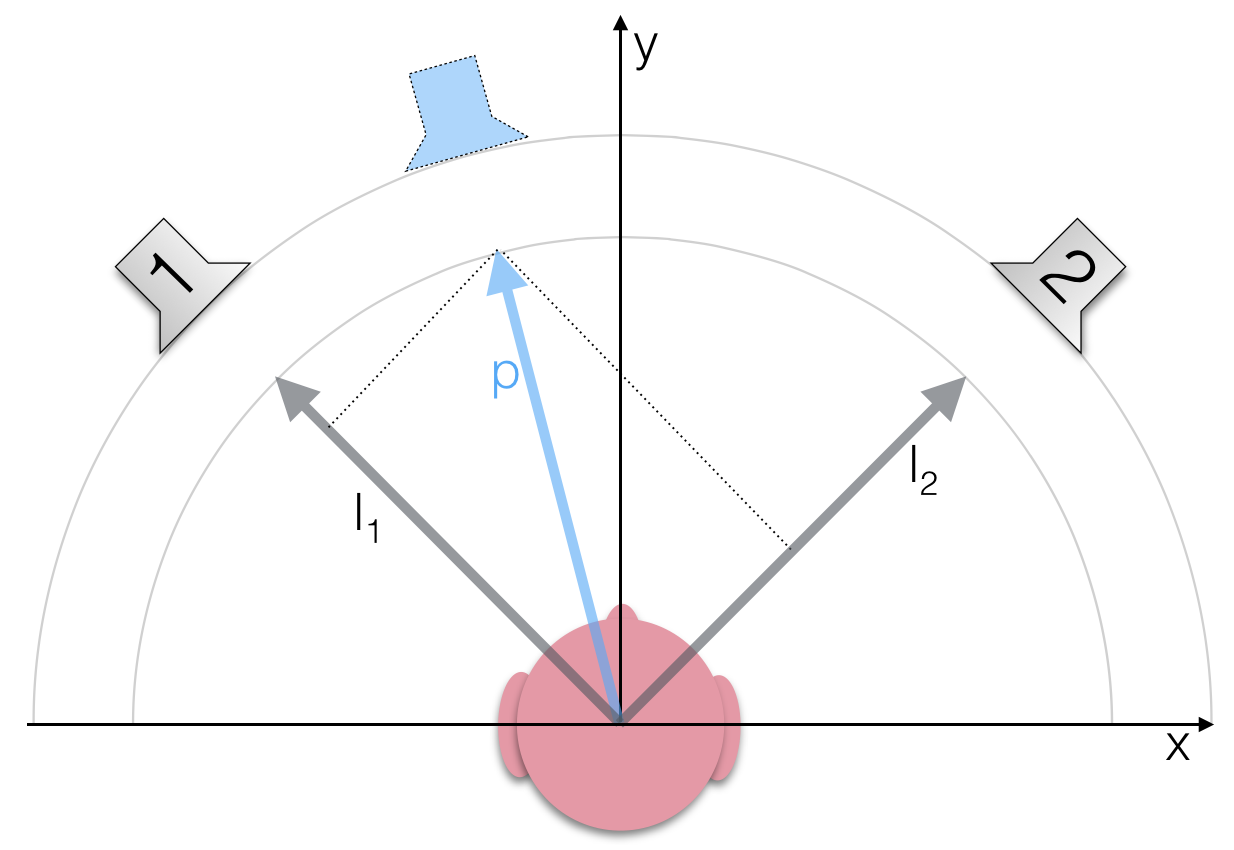}
	\caption[caption]{The goal is to find the vector $p = g_1\cdot l_1 + g_2\cdot l_2$ that points from the listener to the virtual source.} 
	\label{fig_vbap_vectors}
\end{figure}

\begin{equation}
p^T = g \cdot L
\end{equation} 
with $g=\begin{pmatrix} g_{1} & g_{2} & g_{3}\end{pmatrix}$ and $L = \begin{pmatrix} l_{1} & l_{2} & l_{3}\end{pmatrix}^T$ in the three-dimensional case, that is, when three loudspeakers are selected. In the two-dimensional case, $g$ and $L$ have only two components each, since only two loudspeakers are considered at a time. With the inverse $L^{-1}$ of $L$, solving for $g$ yields
\begin{equation} \label{vbap_inverse}
g = p^T \cdot L^{-1}
\end{equation}
The inverse of $L$ exists if the chosen vector base is linearly independent. This always holds, except in border cases where the pair of loudspeakers consists of loudspeakers at diametrically opposing positions or all three loudspeakers from the chosen triplet share their height with the listener. Such cases can be avoided by defining the set of vector bases accordingly. 

Some consequences of this vector approach are worth noting: for virtual sources that share their position with one of the loudspeakers, only the respective loudspeaker will have a non-zero gain, since $p$ will be equal to the loudspeaker's vector $l_i$. Similarly, in the three-dimensional case, if the virtual source is located between two loudspeakers, only these two speakers will have a non-zero gain.

Equation (\ref{vbap_inverse}) is not only used to calculate the gain factor for the selected base, but also to select the base itself: first the gain factors for every base are computed, then the base with no negative gains is selected. Such a base exists if there is a base whose loudspeakers cover the position of the virtual source. In the special case where the virtual source shares its position with one of the loudspeakers, several bases fulfil this condition. This can happen also in the three-dimensional case if the virtual source lays on the arc between two loudspeakers. If several bases are possible candidates, the base with the maximal smallest gain is the preferred choice for reasons of numerical stability. For example, a base with gain factors $g=\begin{pmatrix} 0.2 & 0.3 & 0.3\end{pmatrix}$ is chosen over a base with $g=\begin{pmatrix} 0.1 & 0.9 & 0.9\end{pmatrix}$ since its smallest gain is higher than the smallest gain of the competing base.

In order to generate a moving virtual source with constant perceived loudness, the gain factors $g_i$ have to be normalised. One way to do this is to set the power level to a constant $C$ satisfying $C = g_{norm} \cdot g_{norm}^T$. The unscaled gain vector $g$ from Equation (\ref{vbap_inverse}) can be scaled according to
\begin{equation}
g_{norm} = \sqrt{C} \cdot \frac{g}{\sqrt{g \cdot g^T}}
\end{equation} 
Additionally, the distance among all virtual sources can be controlled so that a source tends to appear closer to the listener for higher values of $C$. However, trying to control the perceived distance through this parameter alone will not yield effective results since a number of psychoacoustical phenomena (e.g. reflections and alterations of the spectrum) need also to be considered. By default, the perceived distance of a virtual source will be the same as the distance to the loudspeakers (Pulkki, 1997).

\subsection{Distance-Based Amplitude Panning}
\emph{Distance-Based Amplitude Panning (DBAP)} is a spatialisation technique first introduced by Lossius et al. in 2009. Unlike VBAP, where the relative directions of loudspeaker and virtual source to the listener are relevant in determining the resulting amplitude, in DBAP the distance between a loudspeaker and the virtual source is the key in determining the individual gain for that loudspeaker. DBAP does not impose restrictions to the loudspeakers' layout or the listener's position, i.e. any number of loudspeakers can be arranged arbitrarily and the listener does not have to be positioned amid them. However, DBAP posits other assumptions and restrictions: the intensity $I$ of a virtual source must be always constant and cannot change with its position. Also, all loudspeakers are active at all times and their individual amplitudes $v_i$ depend on their distance to the virtual source. If the amplitude of the source is also assumed to be $1$, then
\begin{equation} \label{dbap_intensity}
I = \sum_{i=1}^{N} v_i^2 = 1
\end{equation}
holds. The amplitude of a loudspeaker is calculated as
\begin{equation} \label{dbap_amplitude}
v_i = \frac{k}{d_i^a}
\end{equation} where $d_i$ is the Euclidean distance between loudspeaker $i$ and the virtual source, $a$ is a coefficient accounting for the inverse distance law for sound propagating in a free field\footnote{$a=\frac{R}{20\cdot \log_{10} 2}$ with $R = 6 \mathtt{dB}$ in the free field. In closed environments $R$ is around $3\mathtt{dB} \textendash 5 \mathtt{dB}$.} and $k$ is a coefficient that can be calculated by combining Equations (\ref{dbap_intensity}) and (\ref{dbap_amplitude}):
\begin{equation}
1 = \sum_{i=1}^{N} \frac{k^2}{d_i^{2a}} \Leftrightarrow \frac{1}{k^2} = \sum_{i=1}^{N} \frac{1}{d_i^{2a}} \Leftrightarrow k = \frac{1}{\sqrt{\sum_{i=1}^{N} \frac{1}{d_i^{2a}} }}
\end{equation}

A problem with Equation (\ref{dbap_amplitude}) is that it leads to a division by zero if the virtual source is located at the same position as one of the loudspeakers. It can be shown that $\lim_{d_j \to 0}v_i$ is 0 if $i=j$ and 1 otherwise, i.e. only the loudspeaker that shares its position with the virtual source will be active. Fixing this issue by setting the amplitude of one loudspeaker to 1 and all others to 0 might lead to unwanted changes in spatial spread for virtual sources that move across the position of a loudspeaker. A workaround is to introduce a \emph{spatial blur} $r$ when calculating the distance. Now $d_i$ is no longer the Euclidean distance, but
\begin{equation}
d_i = \sqrt{(x_i - x_s)^2 + (y_i - y_s)^2 + (z_i - z_s)^2 + r^2}
\end{equation}
where $(x_i | y_i | z_i)$ and $(x_s | y_s | z_s)$ are the positions of the loudspeaker $i$ and the virtual source $s$, respectively.

An additional step must be taken in order to generate virtual sources that are positioned outside the region covered by the speakers, otherwise localising them becomes a difficult task. In fact, the longer the distance of a virtual source from the loudspeakers, the lower the difference of gains among loudspeakers. Therefore, for virtual sources outside the convex hull described by the geometry of the loudspeakers, the position of the virtual source is set to the closest point inside the convex hull. The distance between this position and the originally intended location of the virtual source can also be used as a parameter for effects such as gain attenuation, Doppler effect and distance dependent reverb (see Figure 13).

\begin{figure}[htb] 
	\centering
  \includegraphics[width=0.5\textwidth ]{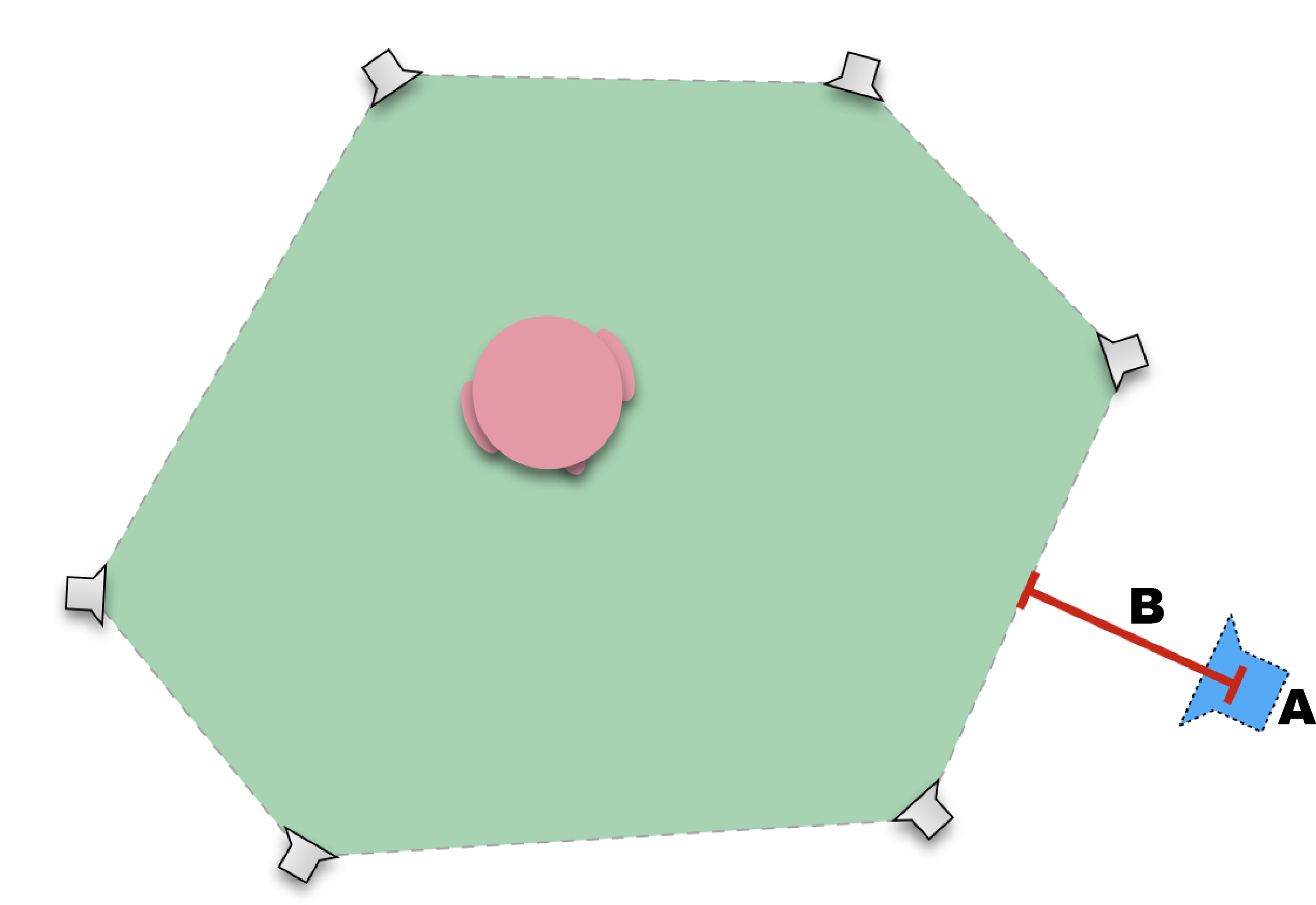}
	\caption[caption]{The polygon describes the convex hull of the loudspeakers. The distance between the virtual source (A) and the convex hull is marked with the letter B.} 
	\label{fig_convex_hull}
\end{figure}

\section{Virtual Acoustic Space}
When using headphones to listen to conventional stereophonic recordings intended for loudspeakers' reproduction, the localisation quality is significantly reduced. Instead of the intended virtual source locations generated via phantom images, the listener perceives the source as though it were inside his/her head. Even with high interaural level/time differences between both channels, the source's location appears closer to the corresponding ear, but not as though it were clearly outside the head.

\emph{Virtual Acoustic Space} or \emph{Virtual Auditory Space} refers to a technique that manipulates sounds in a way that when they are reproduced through headphones the illusion of an external acoustic space is formed. Each virtual sound source can be localised by the listener at the intended position outside the listener's head (Carlile, 2013).

This section evaluates the spatialisation techniques presented for loudspeakers in terms of their compatibility with headphones. Since solely replacing loudspeakers with headphones would generally not yield effective results, alternative approaches to adapt these techniques for use with headphones are described. 

\subsection{Channel-based systems}
\noindent\emph{Stereo.}
At first glance, a traditional stereophonic recording seems to be suitable for playback with headphones because both the recording and the headphones have two channels. In practice, using headphones to listen to sound intended to be reproduced by loudspeakers creates unnatural effects due to both the wide separation of channels in headphones and the lack of acoustic characteristics of the environment. Any virtual source meant to sound outside the listener's head will instead resonate inside it. A number of approaches (Basha et al., 2007; Bauer, 1961; Thomas, 1977) have been introduced to overcome these problems. For example, the method pursued by Basha et al. (2007) to widen the stereo image of a recording consists of the following steps: first the side signal (defined as the difference of the channels $L-R$, where subtraction denotes addition of the inverted signal) is enhanced to increase the side/centre ratio. The next step introduces a crossfeed that simulates the natural crosstalk occurring in a loudspeakers setup where both ears receive signals from both loudspeakers, i.e. the left ear receives the signal from the right loudspeaker after it has travelled through the head and vice versa. Therefore, the crossfeed is accompanied by a low-pass filter to simulate the head's shadow. As a third step, the reflections of the environment are mimicked by a feed-forward delay network. This network adds a delayed and attenuated version of each channel to itself. Again, a low-pass filter is used to account for the stronger absorption of high frequencies by the environment. \\

\noindent\emph{Multi-channel formats.}
Given a multichannel format, a stereo format can be obtained by downmixing. For example in the case of a 5.1 setup, the channels of the front left and back left speakers are simply added to form the left channel of the stereo format (possibly increasing the front channel's amplitude), while the centre channel is resolved by mixing the left and right channels together (with equal amplitude). However, this approach implies that any directional information originally contained in the multichannel signal will be lost. Thus will any spatialisation cues, preventing the listener from detecting the virtual source's position.

The \emph{Dolby Headphone} is a technology that can convert a 5.1 or 7.1 signal into a two channel stereo format without losing directional information. The first step to achieve it is to make use of head-related transfer functions (HRTFs -- see Appendix) to encode each channel's virtual position into the stereo format. The second step mimics the acoustic features of the environment: besides reproducing the direct sound the listener would receive if using loudspeakers, Dolby Headphones also emulate the indirect sound perceived after reflections against the environment's boundaries.

\subsection{Ambisonics} \label{headphone_ambi}
The original conception of Ambisonics only considered playback through loudspeakers, but there are several approaches to obtain a binaural rendering $\begin{bmatrix} L, R \end{bmatrix}$ from the B-format components $W$, $X$, $Y$ and $Z$ -- and possibly additional components in case of higher order ambisonics (Daniel et al., 1998; J\^ot et al., 1998; McKeag \& McGrath, 1996). The general goal is to find a binaural filter matrix $F$ with 
\begin{equation} \label{eq_generalGoal}
\begin{bmatrix}
    L\\
    R\\
\end{bmatrix}
=
 \begin{bmatrix}
   F_{WL} & F_{XL} & F_{YL} & F_{ZL}\\
   F_{WR} & F_{XR} & F_{YR} & F_{ZR}\\
\end{bmatrix}
 \begin{bmatrix}
   W  \\
   X  \\
   Y  \\
   Z  \\
\end{bmatrix}
\end{equation}

A basic approach to obtains a binaural rendering is by using virtual loudspeakers (McKeag \& McGrath, 1996). As a first step, a virtual loudspeaker layout is chosen. Next, the output for these $n$ virtual loudspeakers is computed through a linear combination of the signals of the B-format (W, X, Y and Z) as described in the subsection \textit{Decoding the B-format}. The decoding operation can be summarised as follow:
\begin{equation} \label{eq_decodmatrix}
\begin{bmatrix}
    V_{1}\\
    V_{2}\\
   \vdots\\ 
    V_{n}
\end{bmatrix}
=
 \begin{bmatrix}
   D_{W1} & D_{X1} & D_{Y1} & D_{Z1}\\
   D_{W2} & D_{X2} & D_{Y2} & D_{Z2}\\
   \vdots  & \vdots  & \vdots  & \vdots \\
   D_{Wn} & D_{Xn} & D_{Yn} & D_{Zn}\\
\end{bmatrix}
 \begin{bmatrix}
   W  \\
   X  \\
   Y  \\
   Z  \\
\end{bmatrix}
\end{equation}
where $V_i$ is the output signal of the virtual loudspeaker $i$ and $D_{ki}$ is the corresponding scalar of the decoder $D$. The next step introduces HRTFs to apply the transfer functions from each virtual loudspeaker to each ear:
\begin{equation} \label{eq_hrtfMatrix}
\begin{bmatrix}
    L\\
    R\\
\end{bmatrix}
=
 \begin{bmatrix}
   H_{1L} & H_{2L} & \dots & H_{nL} \\
   H_{1R} & H_{2R} & \dots & H_{nR} \\
\end{bmatrix}
\begin{bmatrix}
    V_{1}\\
    V_{2}\\
   \vdots\\ 
    V_{n}
\end{bmatrix}
\end{equation}
with $H_{ie}$ denoting the HRTF from speaker $i$ to ear $e$. As this equation shows, the final signal for one ear is obtained by summing the signals of all virtual loudspeakers as they would arrive at that ear. Combining equations (\ref{eq_generalGoal}) - (\ref{eq_hrtfMatrix}) yields $F = HD$. Figure 14 illustrates this approach.

\begin{figure}[h]
	\centering
  \includegraphics[width=0.95\textwidth ]{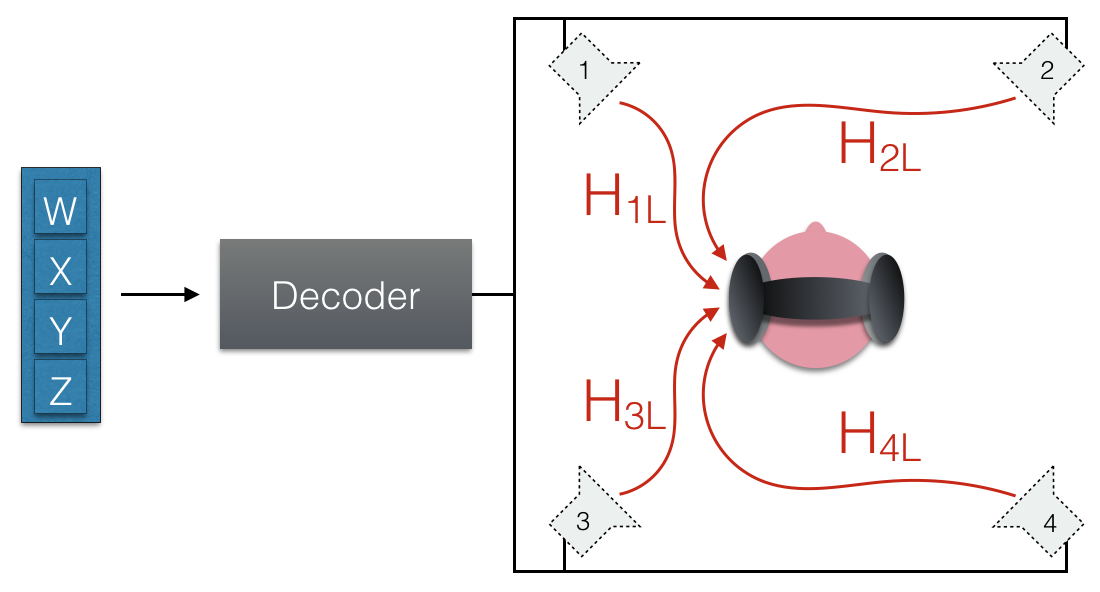}
	\caption[caption]{Obtaining a binaural format via virtual loudspeakers: The decoded signal of each loudspeaker gets processed by the HRTF of the corresponding ear and all signals are added.} 
	\label{fig_ambiVirtuHead}
\end{figure}

An analytical approach to obtain $F$ generates $n$ sources at different positions and uses the relationship between $F$, $H$ and the contributions to the B-format components of the $n$ sources:
\begin{equation} \label{eq_ambiF2}
 \begin{bmatrix}
   H_{1L} & H_{2L} & \dots & H_{nL} \\
   H_{1R} & H_{2R} & \dots & H_{nR} \\
\end{bmatrix}
=
 \begin{bmatrix}
   F_{WL} & F_{XL} & F_{YL} & F_{ZL}\\
   F_{WR} & F_{XR} & F_{YR} & F_{ZR}\\
\end{bmatrix}
 \begin{bmatrix}
   W_{1} & W_{2} & \dots & W_{n} \\
   X_{1} & X_{2} & \dots & X_{n} \\
   Y_{1} & Y_{2} & \dots & Y_{n} \\
   Z_{1} & Z_{2} & \dots & Z_{n} \\
\end{bmatrix}
\end{equation}
where source $i$ contributes the signals $\{W_i, X_i, Y_i, Z_i\}$ and the B-format components are the sums of the contributions, e.g. $W = \sum_i W_i$. The $n$ sources are generated at suitable positions to cover the area surrounding the listener. Since the position of the sources is chosen manually and the encoding of the B-format is determined, the only unknown variables in the above equation are the entries of the matrix $F$. For more than four sources ($n > 4$), Equation (\ref{eq_ambiF2}) becomes mathematically overdetermined and has no general solution. However, the method of least squares is an appropriate way of obtaining $F$.

The assumption of a symmetrical head further simplifies the calculation of $F$ by setting $F_{kR} \coloneqq F_{kL}$. While this assumption does not strictly represent reality, the ears of one individual do not differ as much as the ears of other individuals (J\^ot et al., 1998; McKeag \& McGrath, 1998). 

The approaches presented until this point only work if the user's head remains still while listening. Yet, head rotation is easily implemented if a head tracker is used, as McKeag \& McGrath (1998) did. Alternatively, this information could also be measured by gyro sensors integrated into the headphones, so that a clockwise rotation of the user's head would be equal to a counterclockwise rotation of the acoustic scene. As mentioned in section \ref{manipBformat} (\textit{Manipulating the B-format}), such rotations can be calculated with basic matrix operations. 

In conclusion, it can be said that Ambisonics is very suitable for playback with headphones. In fact, there are several different approaches that account for how to use Ambisonics with headphones instead of a conventional loudspeakers setup. Also, both the traditional B-format consisting of four signals and the higher order variants are viable for use with headphones. The big advantage of Ambisonics -- that its data format is independent of the loudspeakers' layout -- is still maintained when considering headphones and no special steps must be taken to make the recording compatible for them. Only the decoder must be tailored to the setup used for the output, which was anyway the case also with loudspeakers. In the end, the only problem with Ambisonics concerns the use of HRTFs. In fact, HRTFs are not implicitly built in the data format and the acoustic result may vary significantly if the end user does not fit the HRTF database used. However, HRTFs can be changed to adapt to the user's physical features and multiple databases can be offered to the user to choose from.

\subsection{Wave Field Synthesis} \label{headphone_wfs}
Exploiting the physical fundamentals WFS is built on (the Huygens-Fresnel principle) with headphones by directly replacing the loudspeakers' array does not seem feasible, since it requires a certain number of real sound sources to synthesise a wave front with a passable accuracy. Additionally, the headphones' channels are highly separated and do not allow any sound superposition. 

An approach to simulate WFS through headphones was proposed in V\"{o}lk et al. (2008). The basic principle is to replace the array of loudspeakers with virtual loudspeakers (secondary sources). These virtual loudspeakers are then controlled individually -- like real loudspeakers -- to synthesise wavefronts that radiate from a virtual primary source. Instead of exposing the listener to the synthesised wavefronts of real loudspeakers, the user wears headphones and receives the sum of the impulses generated by all virtual loudspeakers that would arrive at the ear positions at the current time. These sound signals must be processed with head-related impulse responses first to encode spatial information about the loudspeakers.

This approach was implemented and tested with a reasonable number of subjects and showed promising results with regard to  localisation accuracy. However, there are still some problems and limitations, the most relevant being that the listener must stand at a fixed position and cannot move around. Therefore, WFS is void of its biggest strength and the result is confined to a narrow space similar to the sweet spot in other approaches. Also, insufficient computing power allows for only a small number of secondary sources (virtual loudspeakers) to be used.

At present, it seems that this approach does not bring any advantages over using HRTFs to encode the position of the primary source. In fact, the use of virtual loudspeakers brings additional issues and limitations without offering any benefits. Despite these drawbacks, Ranjan \& Gun (2015) proposed a hybrid system that uses both loudspeakers and open headphones.\footnote{Unlike closed headphones, open headphones do not isolate the user's ears from the environment, but allow him/her to hear both sounds produced by the headphones and sounds coming from the environment at the same time.} The system consists of an array of 16 real loudspeakers and 32 virtual loudspeakers arranged in front to and around the listener, respectively (see Figure 15). The side and rear scene is realised by encoding spatial information about the virtual loudspeakers through HRTFs, similarly to the approach described above.
 
\begin{figure}[h]
	\centering
  \includegraphics[width=0.6\textwidth ]{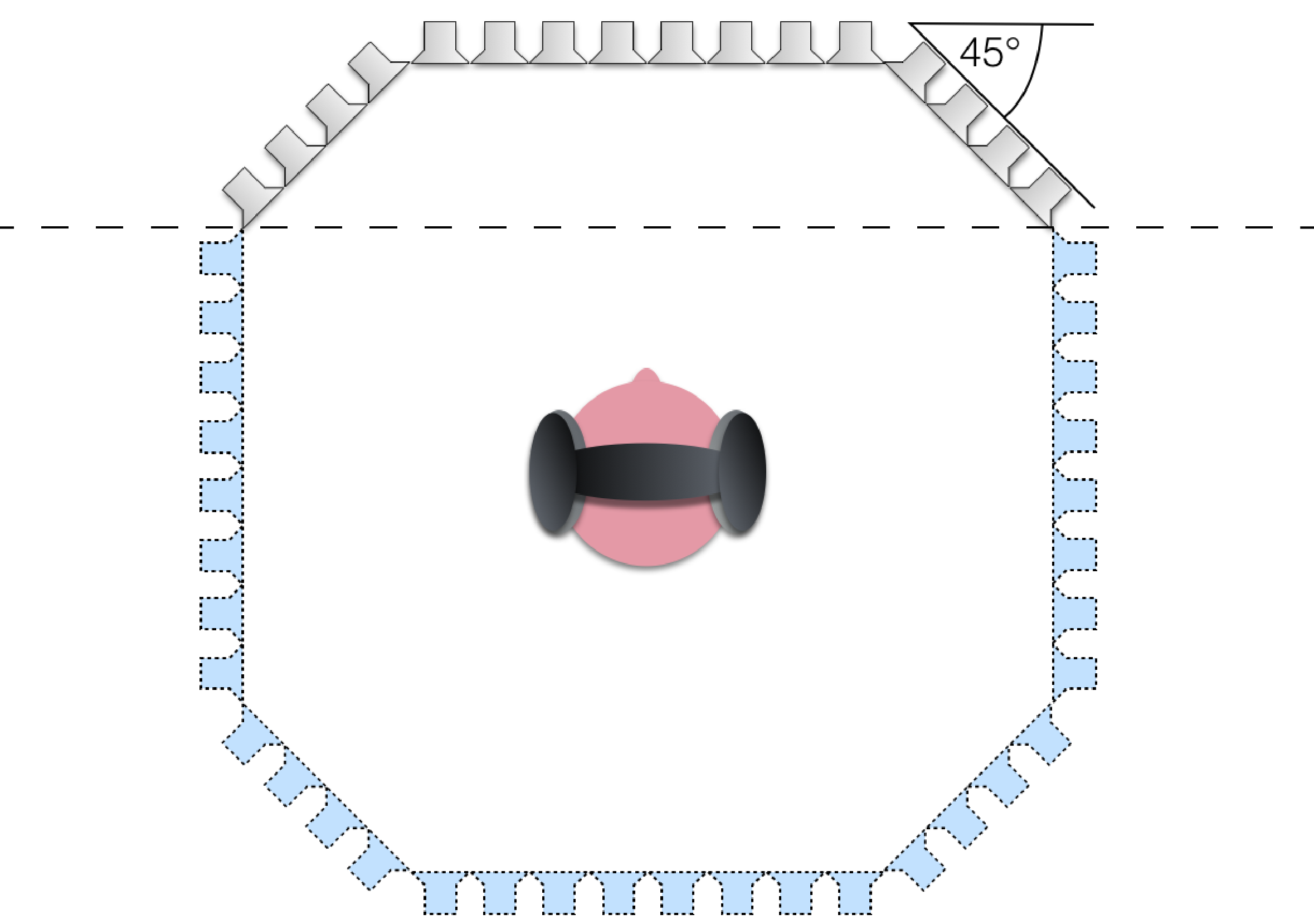}
	\caption[caption]{Hybrid WFS system consisting of real (above the dotted line) and virtual (below the dotted line) loudspeakers proposed in Ranjian \& Gun (2015).} 
	\label{fig_hybridWFS}
\end{figure}

Instead of realising WFS with the use of headphones, there are several approaches that try to go the opposite direction by creating virtual headphones with the use of WFS and crosstalk cancellation (Laumann et al., 2008). These approaches do not technically make use of headphones, but combine a WFS setup with additional features to mimic the characteristics of headphones.

To sum up, attempts to implement the principle of WFS with headphones are promising, but currently there is no system that keeps its advantages when loudspeakers are omitted. 

\subsection{Vector Base Amplitude Panning}
There seem to be no attempts so far at implementing VBAP with headphones. A direct application of the principle behind VBAP, that is, treating headphones like two loudspeakers that are positioned directly at the ears and face the listener, does not seem feasible due to a number of reasons. The first problem is that only two loudspeakers are used to represent the whole horizontal plane, which would result in a maximal error since the loudspeakers are separated by $180^\circ$. A three-dimensional scenario is completely out of the question, since only two loudspeakers are available. Further, the high channel separation in headphones makes it difficult to generate virtual sources that appear outside the head with amplitude panning alone. The mathematical reason for which such an approach is doomed to fail is that the inverse of the matrix $L$ in Equation (\ref{vbap_inverse}) does not exist since the vectors pointing to the loudspeakers are linearly dependent. However, as with other spatialisation techniques, there seem to be alternatives for implementation with headphones. An already explored approach is to just apply the technique as it is, which means replacing the loudspeakers with virtual loudspeakers and sending their outputs to the headphones after processing them with the appropriate HRTF. Since reasonable results are obtained for other approaches and the underlying spatialisation technique is completely abstracted away once the output of the virtual loudspeakers is computed, this strategy should work also with VBAP. In theory, if the transfer functions match the listener's profile perfectly, VBAP should perform with headphones as well as with loudspeakers since the listener would not be able to tell the difference. Of course, in practice this is rather challenging and only achievable if the system is tailored to one person. After all, finding a single HRTF dataset that matches all possible listeners is simply impossible.

Again, a legitimate question would be why not to use HRTFs to encode the positions of the virtual sources instead of introducing the additional complications of VBAP. The reason is that VBAP's HRTF dataset, as well as Ambisonics', may see its complexity be reduced significantly since it needs to contain only a few fixed virtual loudspeakers' directions. In fact, if in a scenario with many virtual images, convolving each of them with individual HRTF operations is computationally expensive, transforming the whole set in a VBAP format and applying the HRTF operation only to the virtual loudspeakers' output is comparatively cheap. Also, the compactness of the HRTF dataset makes easy to quickly adapt it to individual users, as it is the case with Ambisonics. Another possible advantage is that a high number of virtual loudspeakers can be used to minimise the maximal error by reducing the loudspeakers' distances inside a base.  

\subsection{Distance-Based Amplitude Panning}
Like with the other spatialisation approaches, DBAP was designed for loudspeakers surrounding the listener and headphones were not taken into account. Again, treating headphones as though they were two loudspeakers and applying the principle of DBAP would not lead to satisfying results because both headphone channels would emit the same sound but with different gains. Therefore, virtual sources would sound inside instead of outside the listener's head, which is what spatialization techniques aim for.

A rather obvious approach to apply DBAP to a headphone scenario is to simulate loudspeakers through the use of HRTFs, as previously explained. The basic principle remains unchanged: apply the concept of DBAP to create a virtual source, calculate the outputs of virtual loudspeakers and process these outputs with appropriate HRTFs that are computed with the help of a head tracker. However, the strength of DBAP that any number of listeners can be positioned anywhere and thus do not have to keep a fixed distance from the loudspeakers is a matter of some concern. In VBAP the distances between loudspeakers and listener were always the same, then distance could be dismissed altogether and only the relative direction of a virtual loudspeaker had to be encoded via HRTFs. Since this is not the case with DBAP, its implementation with headphones requires additional effort, i.e. both orientation and location of the listeners need to be tracked. Also, far-field HRTF datasets are usually independent of distance, thus the use of HRTF does not automatically encode the distance between a virtual loudspeaker and the listener. In traditional DBAP (with real loudspeakers), this distance was naturally added as the actual distance between the listener and the loudspeakers, whereas with the use of HRTFs, it has to be encoded manually. A possible way of doing this is to scale the loudness of a virtual loudspeaker $i$ based on its distance to the listener with a factor $d_i^a$, where $a$ is a coefficient accounting for the inverse distance law for sound propagating in a free field, as it was introduced in Equation (\ref{dbap_amplitude}). Along with the directional information provided by the HRTF, this can potentially deliver an output that is perceived as coming from a real loudspeaker at the corresponding location. Such a simple way of encoding the distance should only be considered when the listener is in the far field of the loudspeaker. Although Romblom \& Cook (2008) proposed system to overcome difficulties in near-field scenarios, encoding the distance successfully in that case becomes much more challenging, since additional acoustic phenomena -- spectral differences -- and data -- large number of distance calculations -- should be considered. 

DBAP's big strength with loudspeakers is that any number of listeners could be positioned anywhere. Keeping this strength with headphones is costly, both computationally and equipment-wise since each user needs location and orientation trackers. Restricting the number of listeners or their location makes the use of DBAP questionable since there are no real advantages left over other spatialisation techniques. Therefore, although the implementation of DBAP with headphones is not well explored, a quick glance at it reveals more challenges to be solved than there are with other spatialisation techniques, e.g. with Ambisonics. However, if resources are available, this approach can potentially be realised for multiple users at the same time. Also, for a small number of users it is arguably still easier to implement than WFS.

\section{Conclusion} 
Several spatialisation approaches have been surveyed and later evaluated with regard to their compatibility with headphones. In addition to the traditional channel-based techniques that use a pairwise panning approach, Ambisonics, Wave Field Synthesis, Vector Base Amplitude Panning and Distance Based Amplitude Panning were presented. 

With regard to WFS, only a few aspects are compatible with headphones. Strict limitations allow to use the principle behind WFS with headphones only to a certain degree, but, in general, this approach does not offer any practical advantage. Implementation of VBAP and DBAP with headphones is not sufficiently explored yet. A possible approach using virtual loudspeakers and HRTFs, as it was proposed for Ambisonics in several publications, was outlined. In the case of DBAP, some additional concerns were pointed out. All in all, Ambisonics appears to be at present the most compatible spatialisation techniques with headphones, making a HRTF that does not fit the listener's features the only possible point of failure. The algorithm behind Ambisonics does not need to be adapted, only the decoder at the end of the pipeline must be developed with regard to the loudspeakers/headphones' configuration used. This is not a big deal, considering that the decoder needs to be tailored to the specific loudspeakers' configuration anyway. Also, the basic file format of Ambisonics is independent of the loudspeakers/headphones' setup, making its application to different systems -- including headphones -- very practical.

In the end, it is evident that any technique used with headphones makes a heavy use of HRTF datasets. This seemingly universal power of HRTF is particularly problematic, since the HRTF for different subjects can vary significantly, possibly making a perfect simulation for one person nearly useless for others. Nonetheless, an averaged, non-individualised HRTF dataset seems to yield satisfying results for most users. For example, in the case of Ambisonics and VBAP, the HRTF for only a few relative directions is needed, which makes possible to offer the user several datasets to choose from.

One may still wonder why these approaches should be preferred over directly encoding the position of a virtual source with HRTFs, since directly convolving a signal with HRTF yields a perceived virtual source at the corresponding position. The main reason is that, as mentioned above, the HRTF must be usually computed only for a few virtual loudspeakers in a fixed position. Since  convolution is rather costly, this might be a significant upside when a high number of virtual sources and a low number of virtual loudspeakers are used. Also, Ambisonics offers the advantage that the data is inherently stored in a layout-independent format that can be used with loudspeakers even when it was intended for playback through headphones. The same goes for VBAP, since decoding the output for headphones is done by using the loudspeakers' signals as an input.

\newpage
\section*{Appendix}
The ability of humans to identify the location of a sound source in three dimensions even with hearing loss in one ear (Slattery \& Middlebrooks, 1994) hints that binaural cues such as ITD and ILD are not the only cues the human auditory system uses. Before reaching the ears, sound gets shaped in various ways by the listener's body. Reflections from shoulders and torso but also the acoustic shadow of the pinnae, that is, the visible part of the ears, contribute to this. The extent to which different parts of the human body participate in the spatial perception of sound varies depending on the position of the source and the frequency of the sound it emits. The human auditory system uses these data the extract cues for localisation (Musicant \& Butler, 1984). For non-moving sound source and listener, the \emph{Head-Related Transfer Function (HRTF)} $H$ shows the factors involved in sound localisation: 
\begin{equation}
H(r, \theta, \phi, f) = \frac{P(r, \theta, \phi, f)}{P_0(r, f)},
\end{equation}
where $r$, $\theta$ and $\phi$ are the spherical coordinates of the sound source (the origin being at the listener's position), $f$ is the frequency of the sound, $P$ is the sound pressure level at the ear drum and $P_0$ is the hypothetical sound pressure level at the head centre if the listener would not be present (see Figure 16).

Unfortunately, $H$ varies significantly from person to person and is not the same for the left and right ear. Generally speaking, the HRTF is a rather complicated function that cannot be summarized by a simple function of frequency and angle.
For far enough distances between source and listener (about 3 feet) the influence of the distance $r$ on the so-called \emph{far-field HRTF} becomes negligible. With \emph{near-field HRTF} (distance shorter than 3 feet), $r$ is still an important parameter of $H$ (Zhong \& Xie, 2014).

For a given individual, the HRTF can be obtained by measurement. A straightforward way of doing this consists of placing microphones at the subject's ears and capturing the responses to impulses from loudspeakers that are evenly distributed around the subject. The HRTF is the Fourier transform of these so-called \emph{Head-related Impulse Responses (HRIR)}.

Since actual HRTFs of different persons can vary strongly, their performance for a specific subject depends on how closely his/her characteristics match the characteristics of the person used for measurement. Usually, HRTF databases, such as those by IRCAM, MIT and CIPIC, offer different sets measured with different subjects, so that the user can choose the set that fits his/her own HRTF best. So et al. (2010) experimented with an approach that uses clustered non-individualised HRTFs to reduce localisation's errors that naturally occur when using sound stimuli generated by HRTFs that do not fit the listener's personal HRTF. Databases for near-field HRTFs are far less accessible since it is considerably more expensive to obtain them (Zhong \& Xie, 2014).

\newpage
\section*{References}
\begin{list}{}{\setlength\itemindent{-\leftmargin}}
  \item Abramowitz, M. \& Stegun, I. A. (1964). \textit{Handbook of Mathematical Functions: With Formulas, Graphs, and Mathematical Tables}. New York, NY: Dover.
  \item Bamford, J. S. \& Vanderkooy, J. (1995). Ambisonic Sound for Us. In \textit{99th Audio Engineering Society Convention}, New York, NY, October 6-9, n.p.
  \item Berkhout, A. J. (1988). A Holographic Approach to Acoustic Control. \textit{Journal of the Audio Engineering Society, 36(12)}, 977-995.
  \item Basha, S. M. A., Gupta, A. \& Sharma, A. (2007). Stereo Widening System Using Binaural Cues for Headphones. \textit{International Conference on Signal Processing and Communication Systems}, Gold Coast, Australia, December 17-19, n.p.
  \item Bauer, B. B. (1961). Stereophonic Earphones and Binaural Loudspeakers. \textit{The Journal of Audio Engineering Society, 9(2)}, 148-151.
  \item Blauert, J. (1997). \textit{Spatial Hearing: The Psychophysics of Human Sound Localization}. Cambridge, MA: MIT Press.
  \item Daniel, J., Rault, J.-B. \& Polack, J.-D. (1998). Ambisonics Encoding of Other Audio Formats for Multiple Listening Conditions. In \textit{105th Audio Engineering Society Convention}, San Francisco, CA, September 26-29, n.p.
  \item Daniel, J. (2003). Spatial Sound Encoding Including Near Field Effect: Introducing Distance Coding Filters and a Viable, New Ambisonic Format. In \textit{Audio Engineering Society Conference: 23rd International Conference: Signal Processing in Audio Recording and Reproduction}, Helsing\o{}r, Denmark, May 23-25, n.p.
  \item Carlile, S. (2013). \textit{Virtual Auditory Space: Generation and Applications}. Berlin, Germany: Springer.
  \item Dickreiter, M., Dittel, V., Hoeg, W. \& W{\"o}hr, M. (2014). \textit{Handbuch der Tonstudiotechnik}. Berlin, Germany: Walter de Gruyter GmbH \& Co KG.
  \item Elen, R. (2001). Ambisonics: The Surround Alternative. In \textit{Proceedings of the 3rd Annual Surround Conference and Technology Showcase} (pp. 1-4), Beverly Hills, CA, December 7-8, n.p.
  \item Gerzon, M. A. (1980). Practical Periphony: The Reproduction of Full-Sphere Sound. In \textit{65th Audio Engineering Society Convention}, London, UK, February 25-28, n.p.
  \item Gerzon, M. A. \& Barton, G. J. (1984). Ambisonic Surround-Sound Mixing for Multitrack Studios. In \textit{Audio Engineering Society Conference: 2nd International Conference: The Art and Technology of Recording}, Anheim, CA, May 11-14, n.p.
  \item Gerzon, M. A. (1985). Ambisonics in Multichannel Broadcasting and Video. \textit{The Journal of Audio Engineering Society, 33(11)}, 859-871.
  \item Gerzon, M. A. \& Barton, G. J. (1992). Ambisonic Decoders for HDTV. In \textit{92nd Audio Engineering Society Convention}, Vienna, Austria, March 24-27, n.p.
  \item Heller, A., Lee, R. \& Benjamin, E. (2008). Is my Decoder Ambisonic? In \textit{125th Audio Engineering Society Convention}, San Francisco, CA, October 2-5, n.p.
  \item International Telecommunication Union. Recommendation itu-r bs.775-2, 2006.
  \item J\^ot, J.-M., Wardie, S. \& Larcher, V. (1998). Approaches to Binaural Synthesis. In \textit{105th Audio Engineering Society Convention}, San Francisco, CA, September 26-29, n.p.
  \item Kronlachner, M. (2014). \textit{Spatial Transformations for the Alteration of Ambisonic Recordings} (Master's thesis). Graz, Austria: University of Music and Performing Arts.
  \item Laumann, K., Theile, G. \& Fastl, H. (2008). A Virtual Headphone Based on Wave Field Synthesis. In \textit{The Journal of the Acoustical Society of America, 123(5)}, 3515.
  \item Lossius, T., Baltazar, P. \& de la Hogue, T. (2009). Dbap Distance-Based Amplitude Panning. In \textit{International Computer Music Conference}, Montreal, Canada, August 16-21.
  \item McKeag, A. \& McGrath, D. S. (1996). Sound Field Format to Binaural Decoder with Head Tracking. In \textit{Audio Engineering Society Convention 6r}, Lake DSP, Australia, August, n.p.
  \item Malham, D. G. (1998). Spatial Hearing Mechanisms and Sound Reproduction. N.p. Retrieved from \url{https://www.york.ac.uk/inst/mustech/3d_audio/ambis2.htm}
  \item Malham, D. G. (2003). \textit{Space in Music-Music in Space} (Master's thesis). York, UK: University of York.
  \item Martin, G., Woszczyk, W., Corey, J. \& Quesnel, R. (1999). Sound Source Localization in a Five-Channel Surround Sound Reproduction System. In \textit{107th Audio Engineering Society Convention}, New York, NY, September 24-27, n.p.
  \item Musicant, A. D. \& Butler, R. A. (1984). The Influence of Pinnase-Based Spectral Cues on Sound Localization. In \textit{The Journal of the Acoustical Society of America, 75(4)}, 1195-1200.
  \item Penha, R. (2008). Distance Encoding in Ambisonics Using Three Angular Coordinates. In \textit{ Proceedings of the Sound and Music Computing Conference}, Hamburg, Germany, August 31 - September 3, n.p.
  \item Pulkki, V. (1997). Virtual Sound Source Positioning Using Vector Base Amplitude Panning. \textit{Journal of the Audio Engineering Society, 45(6)}, 456-466.
  \item Pulkki, V. \& Karjalainen, M. (2001). Localization of Amplitude-Panned Virtual Sources I: Stereophonic Panning. \textit{The Journal of Audio Engineering Society, 49(9)}, 739-752.
  \item Rabenstein, R. \& Spors, S. (2006). Spatial Aliasing Artifacts Produced by Linear and Circular Loudspeaker Arrays Used for Wave Field Synthesis. In \textit{120th Audio Engineering Society Convention}, Paris, France, May 20-23, n.p.
  \item Ranjan, R. \& Gan, W. S. (2015). A Hybrid Speaker Array-Headphone System for Immersive 3d Audio Reproduction. In \textit{2015 IEEE International Conference on Acoustics, Speech and Signal Processing (ICASSP)} (pp. 1836-1840). Brisbane, Australia, April 19-24.
  \item Schacher, J. C. \& Kocher, P. (2006). Ambisonics Spatialization Tools for Max/MSP. \textit{Omni, 500(1)}, 274-277.
  \item Schnupp, J., Nelken, I. \& King, A. (2011). \textit{Auditory Neuroscience: Making Sense of Sound}. Cambridge, MA: MIT Press.
  \item Slattery, W. \& Middlebrooks, J. C. (1994). Monaural Sound Localization: Acute Versus Chronic Unilateral Impairment. In \textit{Hearing Research, 75(1)}, 38-46.
  \item So, R. H., Ngan, B., Horner, A., Braasch, J. \& Blauert, J. (2010). Toward Orthogonal Non-Individualised Head-Related Transfer Function for Forward and Backward Directional Sound: Cluster Analysis and an Experimental Study. In \textit{Ergonomics, 53(6)}, 767-781.
  \item Theile, G. \& Plenge, G. (1977). Localization of Lateral Phantom Sources. \textit{The Journal of Audio Engineering Society, 25(4)}, 196-200. 
  \item Thomas, M. V. (1977). Improving the Stereo Headphone Sound Image. \textit{The Journal of Audio Engineering Society, 25(7/8)}, 474-478.
  \item Völk, F., Konradl, J. \& Fastl, H. (2008). Simulation of Wave Field Synthesis. In \textit{The Journal of the Acoustical Society of America, 123(5)}, 3159.
  \item Williams, E. G. (1999). \textit{Fourier Acoustics: Sound Radiation and Nearfield Acoustical Holography}. Cambridge, MA: Academic Press.
  \item Zhong, X. \& Xie, B. (2014). Head-Related Transfer Functions and Virtual Auditory Display. In \textit{InTech}. N.p. Retrieved from \url{https://www.intechopen.com/books/soundscape-semiotics-localisation-and-categorisation/head-related-transfer-functions-and-virtual-auditory-display}
\end{list}

\end{document}